\documentclass[runningheads,a4paper]{llncs}

\usepackage{eso-pic}
\AddToShipoutPicture*{
\put(133,160){
\scriptsize
Proceedings of WLPE 2013, {\tt arXiv:1308.2055}, August 2013.
}}

\usepackage{amsmath}
\usepackage{amsfonts}\usepackage{amssymb}
\usepackage{listings}
\usepackage{galois}
\usepackage{url}
\usepackage{tikz}

\newcommand{\ciaopp}{CiaoPP}
\newcommand{\ciao}{Ciao}
\newcommand{\figrule}{\rule{\textwidth}{0.5px}}

\def\imp{\hbox{${\tt \ :\!-\ }$}}

\title{Towards an Abstract Domain for Resource Analysis of Logic Programs using Sized Types}

\titlerunning{Towards Resource Analysis of Logic Programs using Sized Types}

\author{Alejandro Serrano\inst{1} \and
Pedro L\'opez-Garc\'ia\inst{1,2}
\and Manuel Hermenegildo\inst{1,3}
\thanks{The research leading to these results has received funding
from the European Union Seventh Framework Programme (FP7/2007-2013)
under grant agreement no 318337, ENTRA - Whole-Systems Energy
Transparency and the Spanish TIN2012-39391-C04-01 \emph{STRONGSOFT}
project.}
}

\institute{
IMDEA Software Institute
\and Spanish Council for Scientific Research (CSIC)
\and Universidad Polit\'{e}cnica de Madrid (UPM)}

\begin{document}

\label{firstpage}

\maketitle

\begin{abstract}

We present a novel general resource analysis for logic programs
based on sized types.Sized types are representations that
incorporate structural (shape) information and allow expressing both
lower and upper bounds on the size of a set of terms and their
subterms at any position and depth.  They also allow relating the
sizes of terms and subterms occurring at different argument
positions in logic predicates.  Using these sized types, the
resource analysis can infer both lower and upper bounds on the
resources used by all the procedures in a program as functions on
input term (and subterm) sizes, overcoming limitations of existing
analyses and enhancing their precision. Our new resource analysis
has been developed within the abstract interpretation framework, as
an extension of the sized types abstract domain, and has been
integrated into the Ciao preprocessor, CiaoPP.  The abstract
domain operations are integrated with the setting up and solving of
recurrence equations for both, inferring size and resource usage
functions.  We show that the analysis is an improvement over the
previous resource analysis present in CiaoPP and compares well in
power to state of the art systems.
\end{abstract}

\section{Introduction}

\emph{Resource usage analysis} infers the aggregation of some
numerical properties, like memory usage, time spent in computation, or
bytes sent over a wire, throughout the execution of a piece of
code. Such numerical properties are known as \emph{resources}. The
expressions giving the usage of resources are usually given in terms
of the sizes of some input arguments to procedures.

Our starting point is the methodology outlined
by~\cite{granularity,caslog} and~\cite{low-bounds-ilps97},
characterized by the setting up of recurrence equations. In that
methodology, the size analysis is the first of several other analysis
steps that include cardinality analysis (that infers lower and upper
bounds on the number of solutions computed by a predicate), and which
ultimately obtain the resource usage bounds. One drawback of these
proposals, as well as most of their subsequent derivatives, is that
they are only able to cope with size information about subterms in a
very limited way. This is an important limitation, which causes the
analysis to infer trivial bounds for a large class of programs.  For
example, consider a predicate which computes the factorials of a list:

\noindent
\begin{small}
\begin{tabular}{p{6.3cm}p{6cm}}
\begin{lstlisting}
listfact([],   []).
listfact([E|R],[F|FR]) :-
	fact(E, F),
	listfact(R, FR).
	
\end{lstlisting} & 
\begin{lstlisting}
fact(0,1).
fact(N,M) :- N1 is N - 1,
             fact(N1, M1),
             M is N * M1.
\end{lstlisting}
\end{tabular}
\end{small}

\noindent Intuitively, the best bound for the running time of this
program for a list $L$ is $\alpha + \sum_{e \in L} \left( \beta +
  time_{fact}(e) \right)$, where $\alpha$ and $\beta$ are constants
related to the unification and calling costs. But with no further
information, the upper bound for the elements of $L$ must be $\infty$
to be on the safe side, and then the returned overall time bound must
also be $\infty$.

In a previous paper~\cite{sized-types-iclp2013} we focused on a
proposal to improve the size analysis based on \emph{sized
  types}. These sized types are similar to the ones present in
\cite{DBLP:conf/ifl/VasconcelosH03} for functional programs, but our
proposal includes some enhancements to deal with regular types in
logic programs, developing solutions to deal with the additional
features of logic programming such as non-determinism and
backtracking.  While in that paper we already hinted at the fact that
the application of our sized types in resource analysis could result
in considerable improvement, no description was provided of the actual
resource analysis.

This paper is complementary and fills this gap by describing a new
resource usage analysis with two novel aspects. Firstly, it can
\emph{take advantage of the new information contained in sized
  types}. Furthermore, this resource analysis is \emph{fully based on
  abstract interpretation}, i.e., not just the auxiliary analyses but
also the resource analysis itself. This allows us to integrate
resource analysis within the PLAI abstract interpretation
framework~\cite{ai-jlp,inc-fixp-sas} in the \ciaopp{} system, which
brings in features such as \emph{multivariance}, fixpoints, and
assertion-based verification and user interaction for free. We also
perform a performance assessment of the resulting global system.

In Section \ref{sec:overview} we give a high-level view of the
approach.  In the following section we review the abstract
interpretation approach to size analysis using sized types.
Section~\ref{sec:resources} gets deeper into the resource usage
analysis, our main contribution. Experimental results are shown in
Section~\ref{sec:results}. Finally we review some related work and
discuss future directions of our resource analysis work.

\section{Overview of the Approach}
\label{sec:overview}

We give now an overview of our approach to resource usage analysis,
and present the main ideas in our proposal using the classical
\texttt{append/3} predicate as a running example:

\begin{small}
\begin{lstlisting}
append([],    S, S).
append([E|R], S, [E|T]) :- append(R, S, T).
\end{lstlisting}
\end{small}

\noindent
The process starts by performing the regular type analysis present in
the \ciaopp{} system~\cite{eterms-sas02}. In our example, the system
infers that for any call to the predicate \texttt{append(X, Y, Z)}
with \texttt{X} and \texttt{Y} bound to lists of numbers and
\texttt{Z} a free variable, if the call succeeds, then \texttt{Z} also
gets bound to a list of numbers. The set of ``list of numbers'' is
represented by the regular type ``listnum,'' defined as follows:
\begin{lstlisting}
listnum -> [] | .(num, listnum)
\end{lstlisting}

From this regular type definition, sized type schemas are derived. In
our case, the sized type schema $listnum\text{-}s$ is derived from
$listnum$. This schema corresponds to a list that contains a number of
elements between $\alpha$ and $\beta$, and each element is between the
bounds $\gamma$ and $\delta$. It is defined as:
$$listnum\text{-}s \to listnum^{(\alpha,\beta)}(num_{\langle ., 1 \rangle}^{(\gamma, \delta)})$$

From now on, in the examples we will use $ln$ and $n$ instead of
$listnum$ and $num$ for the sake of conciseness. The next phase
involves relating the sized types of the different arguments to the
\texttt{append/3} predicate using recurrence (in)equations.  Let
$size_X$ denote the sized type schema corresponding to argument
\texttt{X} in a call \texttt{append(X, Y, Z)} (created from the
regular type inferred by a previous analysis). We have that $size_X$
denotes $ln^{(\alpha_X, \beta_X)}(n_{\langle ., 1 \rangle}^{(\gamma_X,
  \delta_X)})$.  Similarly, the sized type schema for the output
argument \texttt{Z} is $ln^{(\alpha_Z, \beta_Z)}(n_{\langle ., 1
  \rangle}^{(\gamma_Z, \delta_Z)})$, denoted by $size_Z$. Now, we are
interested in expressing bounds on the length of the output list
\texttt{Z} and the value of its elements as a function of size bounds
for the input lists \texttt{X} and \texttt{Y} (and their
elements). For this purpose, we set up a system of inequations.  For
instance, the inequations that are set up to express a lower bound on
the length of the output argument \texttt{Z}, denoted $\alpha_Z$, as a
function on the size bounds of the input arguments \texttt{X} and
\texttt{Y}, and their subarguments ($\alpha_X,\ \beta_X, \ \gamma_X, \
\delta_X, \ \alpha_Y, \ \beta_Y, \ \gamma_Y$, and $\delta_Y$) are:
\begin{small}
$$
\alpha_Z
\begin{pmatrix}
\alpha_X,\beta_X,\gamma_X,\delta_X, \\
\alpha_Y,\beta_Y,\gamma_Y,\delta_Y
\end{pmatrix}
\geq
\begin{cases}
\alpha_Y & 
\text{if } \alpha_X = 0
\\
1 + \alpha_Z
\begin{pmatrix}
\alpha_X-1,\beta_X-1, \gamma_X,\delta_X, \\
\alpha_Y,\beta_Y,\gamma_Y,\delta_Y
\end{pmatrix} & 
\text{if } \alpha_X > 0
\end{cases}$$
\end{small}

\noindent
Note that in the recurrence inequation set up for the second clause of
\texttt{append/3}, the expression $\alpha_X-1$ (respectively
$\beta_X-1$) represents the size relationship that a lower
(respectively upper) bound on the length of the list in the first
argument of the recursive call to \texttt{append/3} is one unit less
than the length of the first argument in the clause head.

As the number of size variables grows, the set of inequations becomes
too large. Thus, we propose a compact representation. The first change
in our proposal is to write the parameters to size functions directly
as sized types. Now, the parameters to the $\alpha_Z$ function are the
sized type schemas corresponding to the arguments \texttt{X} and
\texttt{Y} of the \texttt{append/3} predicate:
\begin{small}
$$
\alpha_Z
\begin{pmatrix}
ln^{(\alpha_X, \beta_X)}(n_{\langle ., 1 \rangle}^{(\gamma_X, \delta_X)}) \\
ln^{(\alpha_Y, \beta_Y)}(n_{\langle ., 1 \rangle}^{(\gamma_Y, \delta_Y)})
\end{pmatrix}
\geq
\begin{cases}
\alpha_Y &
\text{if } \alpha_X = 0
\\
1 + \alpha_Z
\begin{pmatrix}
ln^{(\alpha_X - 1, \beta_X - 1)}(n_{\langle ., 1 \rangle}^{(\gamma_X, \delta_X)}) \\
ln^{(\alpha_Y, \beta_Y)}(n_{\langle ., 1 \rangle}^{(\gamma_Y, \delta_Y)})
\end{pmatrix} & 
\text{if } \alpha_X > 0
\end{cases}$$
\end{small}

In a second step, we group together all the inequalities of a single
sized type. As we always alternate lower and upper bounds, it is
always possible to distinguish the type of each inequality. We do not
write equalities, so that we do not use the symbol $=$. However, we
always write inequalities of both signs ($\geq$ and $\leq$) for each
size function, since we compute both lower and upper size bounds.
Thus, we use a compact representation $\lessgtr$ for the symbols
$\geq$ and $\leq$ that are always paired. For example, the expression:
$$ln^{(\alpha_X, \beta_X)}(n_{\langle ., 1 \rangle}^{(\gamma_X, \delta_X)})
\lessgtr ln^{(e_1, e_2)}(n_{\langle ., 1 \rangle}^{(e_3, e_4)})$$
represents the conjunction of the following size constraints:
$$\alpha_X \geq e_1, \ \beta_X \leq e_2, \ \gamma_X \geq e_3, \ \delta_X \leq e_4$$

After setting up the corresponding system of inequations for the
output argument \texttt{Z} of \texttt{append/3}, and solving it, we
obtain the following expression:
\begin{small}
$$size_Z\left(
size_X, size_Y
\right) 
\lessgtr
  ln^{( \alpha_X + \alpha_Y, \beta_X + \beta_Y )}
   (n_{\langle ., 1 \rangle}^{(\min(\gamma_X, \gamma_Y), \max(\delta_X, \delta_Y))}) $$
\end{small}
that represents, among others, the relation $\alpha_z \geq \alpha_X +
\alpha_Y$ (resp.  $\beta_z \leq \beta_X + \beta_Y$), expressing that a
lower (resp. upper) bound on the length of the output list \texttt{Z},
denoted $\alpha_z$ (resp. $\beta_z$), is the addition of the lower
(resp. upper) bounds on the lengths of \texttt{X} and \texttt{Y}. It
also represents the relation $\gamma_Z \geq \min(\gamma_X, \gamma_Y)$
(resp. $\delta_Z \leq \max(\delta_X, \delta_Y)$), which expresses that
a lower (resp. upper) bound on the size of the elements of the list
\texttt{Z}, denoted $\gamma_z$ (resp. $\delta_z$), is the minimum
(resp. maximum) of the lower (resp. upper) bounds on the sizes of the
elements of the input lists \texttt{X} and \texttt{Y}.

Resource analysis builds upon the sized type analysis and adds
recurrence equations for each resource we want to analyze. Apart from
that, when considering logic programs, we have to take into account
that they can fail or have multiple solutions when executed, so we
need an auxiliary \emph{cardinality analysis} to get correct results.

Let us focus now on cardinality analysis. Let $s_L$ and $s_U$ denote
lower and upper bounds on the number of solutions respectively that
predicate \texttt{append/3} can generate. Following the program
structure we can infer that:
\begin{small}
$$
\begin{array}{rcl}
s_L\left(
  ln^{(0, 0)}(n_{\langle ., 1 \rangle}^{(\gamma_X, \delta_X)}),
  size_Y
\right) 
& \geq &
  1 \\
s_L\left(
  ln^{(\alpha_X, \beta_X)}(n_{\langle ., 1 \rangle}^{(\gamma_X, \delta_X)}),
  size_Y
\right) 
& \geq &
  s_L\left(
  ln^{(\alpha_X - 1, \beta_X - 1)}(n_{\langle ., 1 \rangle}^{(\gamma_X, \delta_X)}),
  size_Y
\right) 
\end{array}$$
$$
\begin{array}{rcl}
s_U\left(
  ln^{(0, 0)}(n_{\langle ., 1 \rangle}^{(\gamma_X, \delta_X)}),
  size_Y
\right) 
& \leq &
  1 \\
s_U\left(
  ln^{(\alpha_X, \beta_X)}(n_{\langle ., 1 \rangle}^{(\gamma_X, \delta_X)}),
  size_Y
\right) 
& \leq &
  s_U\left(
  ln^{(\alpha_X - 1, \beta_X - 1)}(n_{\langle ., 1 \rangle}^{(\gamma_X, \delta_X)}),
  size_Y
\right) 
\end{array}$$
\end{small}
The solution to these inequations is $(s_L, s_U) = (1,1)$, so we have
inferred that \texttt{append/3} generates at least (and at most) one
solution.  Thus, it behaves like a function.  When setting up the
equations, we have used our knowledge that \texttt{append/3} cannot
fail when given lists as arguments. If not, the lower bound in the
number of solutions would be 0.

Now we move forward to analyzing the number of resolution steps
performed by a call to \texttt{append/3} (we will only focus on upper
bounds, $r_u$, for brevity). For the first clause, we know that only
one resolution step is needed, so:
\begin{small}
$$r_U\left(
  ln^{(0, 0)}(n_{\langle ., 1 \rangle}^{(\gamma_X, \delta_X)}),
  ln^{(\alpha_Y, \beta_Y)}(n_{\langle ., 1 \rangle}^{(\gamma_Y,
    \delta_Y)}) \right) \leq 1 $$
\end{small}

\noindent The second clause performs one
resolution step plus all the resolution steps performed by all possible
backtrackings over the call in the body of the clause. This number of
possible backtrackings is bounded by the number of solutions of the
predicate. So the equation reads:
\begin{small}
$$\begin{array}{rclcl}
  r_U\left(
    ln^{(\alpha_X, \beta_X)}(n_{\langle ., 1 \rangle}^{(\gamma_X, \delta_X)}),
    size_Y
  \right) 
  & \leq & 1 & + & 
  s_U\left(
  ln^{(\alpha_X - 1, \beta_X - 1)}(n_{\langle ., 1 \rangle}^{(\gamma_X, \delta_X)}),
  size_Y
  \right)
\\ & & & \times &
  r_U\left(
  ln^{(\alpha_X - 1, \beta_X - 1)}(n_{\langle ., 1 \rangle}^{(\gamma_X, \delta_X)}),
  size_Y
  \right)
\\ & = & 1 & + & r_U\left(
  ln^{(\alpha_X - 1, \beta_X - 1)}(n_{\langle ., 1 \rangle}^{(\gamma_X, \delta_X)}),
  size_Y
  \right)
\end{array}$$
\end{small}

\noindent Solving these equations we infer that an upper bound on the number of
resolution steps is the (upper bound on the length) of the input list
\texttt{X} plus one. This is expressed as:
\begin{small}
$$r_U\left(
    ln^{(\alpha_X, \beta_X)}(n_{\langle ., 1 \rangle}^{(\gamma_X, \delta_X)}),
    ln^{(\alpha_Y, \beta_Y)}(n_{\langle ., 1 \rangle}^{(\gamma_Y, \delta_Y)}) \right)
  \leq \beta_X + 1$$
\end{small}

\section{Sized Types Review}
\label{sec:sizedtypes}

As shown in the \texttt{append} example, the (bound) variables that we
relate in our inequations come from sized types, which are ultimately
derived from the regular types previously inferred for the
program. Among several representations of regular types used in the
literature, we use one based on \emph{regular term grammars},
equivalent to~\cite{Dart-Zobel} but with some adaptations.  A
\emph{type term} is either a \emph{base type} $\alpha_i$ (taken from a
finite set), a \emph{type symbol} $\tau_i$ (taken from an infinite
set), or a term of the form $f(\phi_1, \dots, \phi_n)$, where $f$ is a
$n$-ary function symbol (taken from an infinite set) and $\phi_1,
\dots, \phi_n$ are \emph{type terms}.  A \emph{type rule} has the form
$\tau \to \phi$, where $\tau$ is a \emph{type symbol} and $\phi$ a
\emph{type term}. A \emph{regular term grammar} $\Upsilon$ is a set of
\emph{type rules}.

To devise the abstract domain we focus specifically on the generic
\textsc{and-or} trees procedure of
\cite{DBLP:journals/jlp/Bruynooghe91}, with the optimizations of
\cite{ai-jlp}. This procedure is \emph{generic} and goal dependent: it
takes as input a pair $(L,\lambda_c)$ representing a predicate along
with an abstraction of the call patterns (in the chosen \emph{abstract
  domain}) and produces an abstraction $\lambda_o$ which
overapproximates the possible outputs. This procedure is the basis of
the PLAI abstract analyzer present in
\ciaopp~\cite{hermenegildo11:ciao-design-tplp}, where we have
integrated an implementation of the proposed size analysis.

The formal concept of \emph{sized type} is an abstraction of a set of
Herbrand terms which are a subset of some regular type $\tau$ and meet
some lower- and upper-bound size constraints on the number of
\emph{type rule applications}. A grammar for the new sized types
follows:

\medskip
\noindent
\figrule

\begin{tabular}{rclr}
\emph{sized-type} & $::=$  & $\alpha^{bounds}$ & $\alpha$ base type \\
                  & $|$    & $\tau^{bounds}(\textit{sized-args})$ & $\tau$ recursive type symbol \\
                  & $|$    & $\tau(\textit{sized-args})$ & $\tau$ non-recursive type symbol \\
\emph{bounds}     & $::=$  & $nob$ \ $|$ \ $(n, m)$  & $n,m \in \mathbb{N}, m \geq n$ \\
\emph{sized-args} & $::=$  & $\epsilon$  \ $|$ \ \emph{sized-arg}, \ \emph{sized-args}   \\
\emph{sized-arg}  & $::=$  & $\textit{sized-type}_{position}$ \\
\emph{position}   & $::=$  & $\epsilon$ \ $|$ \ $\langle f, n \rangle$ & $f$ functor, $0 \leq n \leq$ arity of $f$ \\
\end{tabular}

\noindent\figrule

\medskip
\noindent However, in our abstract domain we need to refer to sets of
sized types which satisfy certain constraints on their bounds. For
that purpose, we introduce \emph{sized type schemas}: a schema is just
a sized type with variables in bound positions, along with a set of
constraints over those variables. We call such variables \emph{bound
  variables}. We will denote $sized(\tau)$ the sized type schema
corresponding to a regular type $\tau$ where all the bound variables
are fresh.

\

The full abstract domain is an extension of sized type schemas to
several predicate variables.  Each abstract element is a triple
$\left\langle t, d, r \right\rangle$ such that:
\begin{enumerate}
\item $t$ is a set of $v \to (sized(\tau),c)$, where $v$ is a
  variable, $\tau$ its regular type and $c$ is its
  classification. Subgoal variables can be classified as
  \emph{output}, \emph{relevant}, or \emph{irrelevant}. Variables
  appearing in the clause body but not in the head are classified as
  \emph{clausal};
\item $d$ (the \emph{domain}) is a set of constraints over the
  relevant variables;
\item $r$ (the \emph{relations}) is a set of relations among bound
  variables.
\end{enumerate}

For example, the final abstract elements corresponding to the clauses
of the \verb"listfact" example can be found below. The equations have
already been normalized into their simplest form for conciseness:
\begin{small}
$$\lambda'_1 = \left\langle
\begin{array}{c}
\left\{ L \to (ln^{(\alpha_1,\beta_1)}(n^{(\gamma_1,\delta_1)}), rel.),
       FL \to (ln^{(\alpha_2,\beta_2)}(n^{(\gamma_2,\delta_2)}), out.) \right\} \\
\{ \alpha_1 = 1, \beta_1 = 1 \},
\{ ln^{(\alpha_2,\beta_2)}(n^{(\gamma_2,\delta_2)}) \lessgtr ln^{(1,1)}(n^{nob}) \}
\end{array}
\right\rangle$$

$$\lambda'_{2} = \left\langle
\begin{array}{c}
\left\{ 
\begin{array}{c}
L \to (ln^{(\alpha_1,\beta_1)}(n^{(\gamma_1,\delta_1)}), rel.),
FL \to (ln^{(\alpha_2,\beta_2)}(n^{(\gamma_2,\delta_2)}), out.), \\
E \to (n^{(\gamma_3,\delta_3)}, cl.),
R \to (ln^{(\alpha_4,\beta_4)}(n^{(\gamma_4,\delta_4)}), cl.), \\
F \to (n^{(\gamma_5,\delta_5)}, cl.),
FR \to (ln^{(\alpha_6,\beta_6)}(n^{(\gamma_6,\delta_6)}), cl.)
\end{array}
\right\} \\
\{ \alpha_1 > 0, \beta_1 > 0 \}, \\
\left\{
\begin{array}{c}
ln^{(\alpha_2, \beta_2)}(n^{(\gamma_2, \delta_2)}) 
 \lessgtr
 ln^{(\alpha' + 1, \beta' + 1)}(n^{(\min(\gamma_1 !, \gamma'), \max(\delta_1 !, \delta')}) \\
ln^{(\alpha', \beta'}(n^{(\gamma', \delta')}) 
 \lessgtr 
 factlist\left(ln^{(\alpha_1 - 1, \beta_1 - 1)}(n^{(\gamma_1, \delta_1)}) \right)
\end{array}
\right\}
\end{array}
\right\rangle$$
\end{small}

\section{The Resources Abstract Domain}
\label{sec:resources}

We take advantage of the added power of sized types to develop a
better resource analysis which infers upper and lower bounds on the
amount of resources used by each predicate as a function of the sized
type schemas of the input arguments (which encode the sizes of the
terms and subterms appearing in such input arguments).  For this
reason, the novel abstract domain for resource analysis that we have
developed is tightly integrated with the sized types abstract domain.
 
Following~\cite{resource-iclp07}, we account for two places where the
resource usage can be abstracted:
\begin{itemize}
\item When entering a clause: some resources may be needed during
  unification of the call (subgoal) and the clause head, the
  preparation of entering that clause, and any work done when all the
  literals of the clause have been processed. This cost, dependent on
  the head, is called \emph{head cost}, $\beta$.
\item Before calling a literal: some resources may be used to prepare
  a call to a body literal (e.g., constructing the actual
  arguments). The amount of these resources is known as \emph{literal
    cost} and is represented by $\delta$.
\end{itemize}

\

We first consider the case of estimating upper bounds on resource
usages. For simplicity, assume also that we deal with predicates
having a behavior that is close to functional or imperative programs,
i.e., that are deterministic and do not fail. Then, we can bound the
resource consumption of a clause \\ \centerline{$C \equiv p(\bar{x})
  \imp q_1(\bar{x}_1), \dots, q_n(\bar{x}_n)$,} denoted $r_{U,clause}$
using the formula:
$$r_{U,clause}(C)
  \leq \beta(p(\bar{x})) + \textstyle\sum_{i=1}^n
    \left( \delta(q_i(\bar{x}_i)) + r_{U,pred}(q_i(\bar{x}_i)) \right)$$

    As in sized type analysis, the sizes of some input arguments may
    be explicitly computed, or, otherwise, we express them by using a
    generic expression, giving rise (in the case of recursive clauses)
    to a recurrence equation that we need to solve in order to find
    closed form resource usage functions.

    The resource usage of a predicate, $r_{U,pred}$, depending on its
    input data sizes, is obtained from the resource usage of the
    clauses defining it, by taking the maximum of the equations that
    meet the constraints on the input data sizes (i.e., have the same
    domain).

However, in logic programming we have two extra features to take care of:
\begin{itemize}
\item We may execute a literal more than once on backtracking. To
  bound the number of times a literal is executed, we need to know the
  \emph{number of solutions} each literal (to its left) can
  generate. Using that information, the number of times a literal is
  executed is at most the product of the upper bound on the number of
  solutions, $s_U$, of all the previous literals in the clause. We get
  then the relation:
\begin{small}
$$
\begin{array}{l}
r_{U,clause}\left( p(\bar{x}) \imp q_1(\bar{x}_1), \dots, q_n(\bar{x}_n) \right) \\
\quad  \leq \beta(p(\bar{x})) + \sum_{i=1}^n 
    \left( \prod_{j = 1}^{i-1}s_{pred}(q_j(\bar{x}_j)) \right)
    \left( \delta(q_i(\bar{x}_i)) + r_{U,pred}(q_i(\bar{x}_i)) \right)
\end{array}
$$
\end{small}
\item Also, in logic programming more than one clause may unify with a
  given subgoal. In that case it is incorrect to take the maximum of
  the resource usages of clauses when setting up the recurrence
  equations. A correct solution is to take the sum of every set of
  equations with a common domain, but the bound becomes then very
  rough. Finer-grained possibilities can be considered by using
  different \emph{aggregation} procedures per resource.
\end{itemize}

Lower bounds analysis is similar, but needs to take into account the
possibility of failure, which stops clause execution and forces
backtracking. Basically, no resource usage should be added beyond the
point where failure may happen.  For this reason, in our
implementation of the abstract domain we use the non-failure analysis
already present in \ciaopp. Also, the aggregation of clauses with a
common domain must be different to that used in the upper bounds
case. The simplest solution is to just take the minimum of the
clauses. However, this again leads to very rough bounds. We will
discuss lower bound aggregation later.

\paragraph{\textbf{Cardinality Analysis}}
We have already discussed why cardinality analysis (which estimates
bounds on the number of solutions) is instrumental in resource
analysis of logic programs. We can consider the number of solutions as
another resource, but, due to its importance, we treat it separately.

An upper bound on the number of solutions of a single clause could be
gathered by multiplying the number of solutions of all possible
clauses:
$$s_{U,clause}\left( p(\bar{x}) \imp q_1(\bar{x}_1), \dots,
  q_n(\bar{x}_n) \right) = \textstyle\prod_{i=1}^n
s_{U,pred}(q_i(\bar{x_i}))$$ For aggregation we need to add the
equations with a common domain, to get a recurrence equation
system. These equations will be solved later to get a closed form
function giving an upper bound on the number of solutions.

It is important to remark that many improvements can be added to this
simple cardinality analysis to make it more precise. Some of them are
discussed in~\cite{caslog}, like maintaining separate bounds for the
relation defined by the predicate and the number of solutions for a
particular input, or dealing with mutually exclusive clauses by
performing the $\max$ operation, instead of the addition operation
when aggregating. However, our focus here is the definition of an
abstract domain, and see whether a simple definition produces
comparable results for the resource usage analysis.

One of the improvements we decided to include is the use of the
determinacy analysis present in \ciaopp~\cite{determinacy-ngc09}. If
such analysis infers that a predicate is deterministic, we can safely
set the upper bound for the number of solutions to 1, avoiding the
setting up of recurrence equations.

In the case of lower bounds, we need to know for each clause whether
it may fail or not. For that reason we use the non-failure analysis
already present in \ciaopp~\cite{nfplai-flops04}. In case of a
possible failure, the lower bound on the number of solutions is set to
0.

\paragraph{\textbf{The Abstract Elements}}
The abstract elements are derived from sized type analysis by adding
some extra components. In particular:
\begin{enumerate}
\item The \emph{current variable for solutions}, and \emph{current
    variable for each resource}.
\item A boolean element for telling whether we have already found a
  failing literal.
\item Information about non-failure analysis, coming from its
  abstract domain.
\item Information about determinacy analysis, coming from its
  abstract domain.
\end{enumerate}
We will denote the abstract elements by
$$\langle (s_L, s_U), v_{resources}, failed?, d, r, nf, det \rangle$$
where $(s_L, s_U)$ are the lower and upper bound variables for the
number of solutions, $v_{resources}$ is a set of pairs $(r_L, r_U)$
giving the lower and upper bound variables for each resource,
$failed?$ is a boolean element (either {\tt true} or {\tt false}), $d$
and $r$ are defined as in the sized type abstract domain, and $nf$ and
$det$ can take the values {\tt not\_fails}/{\tt fails} and {\tt
  non\_det}/{\tt is\_det} respectively.

In this analysis we assume that we are given the definition of a set
of resources, which are fixed throughout the whole analysis
process. We have already mentioned three operations, but we need an
extra one for having a complete algorithm. For each resource $r$ we
have:
\begin{itemize}
\item Its head cost, $\beta_r$, which takes a clause head as parameter;
\item Its literal cost, $\delta_r$, which takes a literal as parameter;
\item Its aggregation procedure, $\Gamma_r$, which takes the equations
  for each of the clauses and creates a new set of recurrence
  equations from them;
\item The default upper $\bot_{r,U}$ and lower $\bot_{r,L}$ bound on
      resource usage. 
\end{itemize}

To better understand how the domain works, we will continue with the
analysis of the \verb"listfact" predicate that we started in the
previous section. We assume that the only resource to be analyzed is
the ``number of steps,'' so that we use the following values for the
parameters of the resource analysis:
$$\beta = 1, \quad \delta = 0, \quad \Gamma_r = +,
\quad (\bot_L, \bot_U) = (0,0)$$

\paragraph{\textbf{$\sqsubseteq$, $\sqcup$ and $\bot$}}
We do not have a decidable definition for $\sqsubseteq$ or $\sqcup$,
because there is no general algorithm for checking the inclusion or
union of sets of integers defined by recurrence relations. Instead, we
just check whether one set of inequations is a subset of another one,
up to variable renaming, or perform a syntactic union of the
inequations. This is enough for having a correct analysis.

For $\bot$ we first generate new variables for each of the resources
and the solution. Then, we add relations between them and the default
cost for each resource. For an unknown predicate, the number of
solutions could be any natural number, so we take it as $[0, \infty)$.
We also assume that the predicate may fail.

As mentioned before, the components for non-failure and determinacy
come from the abstract domains for those analyses.

For example, the bottom element for the ``number of steps'' resource will be
(where $\bot_{nf}$ and $\bot_{det}$ are the bottom elements in the
non-failure and determinacy domains respectively):
$$\langle (s_L, s_U), \{ (n_L, n_U) \}, \texttt{true}, \emptyset, 
\{ (s_L, s_U) \lessgtr (0, \infty), (n_L, n_U) \lessgtr (0,0) \},
\bot_{nf}, \bot_{det} \rangle$$

\paragraph{\textbf{$\lambda_{call}$ to $\beta_{entry}$}}
In this operation we need to create the initial structures for
handling the bounds on the number of solutions and resources. This
implies the generation of fresh variables for each of them, and
setting them to their initial values. In the case of the number of
solutions, the initial value is 1 (which is the number of solutions
generated by a fact, and also the neutral element of the product which
appears in the corresponding formula). For a resource $r$, the initial
value is exactly $\beta_r$.

The addition of constraints over sized types when the head arguments
are partially instantiated is inherited from the sized types domain.
Finally, for the $failed?$ component, we should start with value
\texttt{false}, as no literal has been executed yet, so it cannot
fail.

\

In the \verb"listfact" example, the entry substitutions are:
\begin{small}
$$\beta_{entry,1} = 
\left\langle
\begin{array}{c}
(s_{L,1,1}, s_{U,1,1}), \{ (n_{L,1,1}, n_{U,1,1}) \}, \texttt{false},
\{ \alpha_1 = 0, \beta_1 = 0 \}, \\
\{ (s_{L,1,1}, s_{U,1,1}) \lessgtr (1,1), (n_{L,1,1}, n_{U,1,1}) \lessgtr (1,1) \},
\texttt{not\_fails}, \texttt{is\_det}
\end{array}
\right\rangle$$
$$\beta_{entry,2} = 
\left\langle
\begin{array}{c}
(s_{L,2,1}, s_{U,2,1}), \{ (n_{L,2,1}, n_{U,2,1}) \}, \texttt{false},
\{ \alpha_1 > 0, \beta_1 > 0 \}, \\
\{ (s_{L,2,1}, s_{U,2,1}) \lessgtr (1,1), (n_{L,2,1}, n_{U,2,1}) \lessgtr (1,1) \},
\texttt{not\_fails}, \texttt{is\_det}
\end{array}
\right\rangle$$
\end{small}

\paragraph{\textbf{The Extend Operation}}
In the \emph{extend} operation we get both the current abstract
substitution and the abstract substitution coming from the literal
call. We need to update several components of the abstract
element. First of all, we need to include a call to the function
giving the number of solutions and the resource usage from the called
literal.

Afterwards, we need to generate new variables for the number of
solutions and resources, which will hold the bounds for the clause up
to that point. New relations must be added to the abstract element to
give a value to those new variables:
\begin{itemize}
\item For the number of solutions, let $s_{U,c}$ be the new upper
  bound variable, $s_{U,p}$ the previous variable defining an upper
  bound on the number of solutions, and $s_{U,\lambda}$ an upper bound
  on the number of solutions for the subgoal. Then we need to include
  an assignment: $s_{U,c} \leq s_{U,p} \times s_{U,\lambda}$.

  In the case of lower bound analysis, there are two phases. First of
  all, we check whether the called literal can fail, looking at the
  output of the non-failure analysis. If it is possible for it to
  fail, we update the $failed?$ component of the abstract element to
  {\tt true}. If after this the $failed?$ component is still {\tt
    false} (meaning that neither this literal nor any of the previous
  ones may fail) we include a relation similar to the one for upper
  bound case: $s_{L,c} \geq s_{L,p} \times s_{L,\lambda}$. Otherwise,
  we include the relation $s_{L,c} \geq 0$, because failing predicates
  produce no solutions.

\item The approach for resources is similar. Let $r_{U,c}$ be the new
  upper bound variable, $r_{U,p}$ the previous variable defining an
  upper bound on that resource and $r_{U,\lambda}$ an upper bound on
  resources from the analysis of the literal. The relation added in
  this case is $r_{U,c} \leq r_{U,p} + s_{U,p} \times \left( \delta +
    r_{U,\lambda} \right)$.

  For lower bounds, we have already updated the $failed?$ component,
  so we only have to work in consequence. If the component is still
  {\tt false}, we add a new relation similar to the one for upper
  bounds. If it is {\tt true}, it means that failure may happen at
  some point, so we do not have to add that resource any more. Thus
  the relation to be included would be $r_{L,c} \geq r_{L,p}.$
\end{itemize}

\vspace{-0.1cm}

\noindent
In our example, consider the extension of \verb"listfact" after
performing the analysis of the \verb"fact" literal, whose resource
components of the abstract element will be:
\begin{small}
$$\left\langle
\begin{array}{c}
(s_{L}, s_{U}), \{ (n_{L}, n_{U}) \}, \texttt{false}, \{ \alpha, \beta \geq 0 \} \\
\{ (s_{L}, s_{U}) \lessgtr (1,1), (n_{L}, n_{U}) \lessgtr (\alpha,\beta) \}, 
\texttt{not\_fails}, \texttt{is\_det}
\end{array}
\right\rangle$$
\end{small}
As this literal is known not to fail, we do not change the value of
the $failed?$ component of our abstract element for the second
clause. That means that it is still {\tt false}, so we add complete
calls:
\begin{small}
$$\beta_{entry,2} = 
\left\langle
\begin{array}{c}
(s_{L,2,2}, s_{U,2,2}), \{ (n_{L,2,2}, n_{U,2,2}) \}, \texttt{false}, \{ \dots \} \\
\left\{
\begin{array}{c}
 \dots, \\
 (s_{L,2,2}, s_{U,2,2}) \lessgtr (1 \times s_{L,2,1}, 1 \times s_{U,2,1}), \\
(n_{L,2,2}, n_{U,2,2}) \lessgtr (\gamma_1 + n_{L,2,1}, \delta_1 + n_{U,2,1})
\end{array} \right\}, \\
\texttt{not\_fails}, \texttt{is\_det}
\end{array}
\right\rangle$$
\end{small}

\vspace{-0.3cm}

\paragraph{\textbf{$\beta_{exit}$ to $\lambda'$}}
After performing all the extend operations, the variables appearing in
the number of solutions and resources positions will hold the correct
value for their respective numerical properties. As we did with sized
types, we follow now a normalization step, based on ~\cite{caslog}: we
replace each variable appearing in a expression with its definition in
terms of other variables, in reverse topological order, starting from
the desired variables. Following this process, we should reach the
variables in the sized types of the input parameters in the clause
head.

\

Going back to our \verb"listfact" example, the final substitutions
would be:
\begin{small}
$$\lambda'_{1} = 
\left\langle
\begin{array}{c}
(s_{L,1,1}, s_{U,1,1}), \{ (n_{L,1,1}, n_{U,1,1}) \}, \texttt{false},
\{ \alpha_1 = 0, \beta_1 = 0 \}, \\
\{ (s_{L,1,1}, s_{U,1,1}) \lessgtr (1,1), (n_{L,1,1}, n_{U,1,1}) \lessgtr (1,1) \},
\texttt{not\_fails}, \texttt{is\_det}
\end{array}
\right\rangle$$
$$\lambda'_{entry,2} = 
\left\langle
\begin{array}{c}
(s_{L,2,3}, s_{U,2,3}), \{ (n_{L,2,3}, n_{U,2,3}) \}, \texttt{false},
\{ \alpha_1 > 0, \beta_1 > 0 \}, \\
\left\{
\begin{array}{c}
s_{L,2,3} \geq 1 \times listfact_{sol.,L}(ln^{(\alpha_1 - 1,\beta_1 - 1)}(n^{(\gamma_1,\delta_1)})), \\
s_{U,2,3} \leq 1 \times listfact_{sol.,U}(ln^{(\alpha_1 - 1,\beta_1 - 1)}(n^{(\gamma_1,\delta_1)})), \\
n_{L,2,3} \geq \gamma_1 + listfact_{no.\,steps,L}(ln^{(\alpha_1 - 1,\beta_1 - 1)}(n^{(\gamma_1,\delta_1)})), \\
n_{U,2,3} \leq \delta_1 + listfact_{no.\,steps,L}(ln^{(\alpha_1 - 1,\beta_1 - 1)}(n^{(\gamma_1,\delta_1)}))
\end{array}
\right\}, \\
\texttt{not\_fails}, \texttt{is\_det}
\end{array}
\right\rangle$$
\end{small}

\paragraph{\textbf{Widening $\nabla$ and Closed Forms}}
As mentioned before, in contrast to previous cost analyses, at this
point we bring in the possibility of different aggregation
operators. Thus, when we have the equations, we need to pass them to
each of the corresponding $\Gamma_r$ per each resource $r$ to get the
final equations.

This process can be further refined in the case of solution analysis,
using the information from the non-failure and determinacy analyses.
If the final output of the non-failure analysis is {\tt fails}, we
know that the only correct lower bound is 0. So we can just assign the
relation $s_L \geq 0$ without further recurrence relation setting.
Conversely, if the final output of the determinacy analysis is {\tt
  is\_det}, we can safely set the relation $s_U \leq 1$, because at
most one solution will be produced in each case. Furthermore, we can
refine the lower bound on the number of solutions with the minimum
between the current bound and 1.

In the example analyzed above there was an implicit assumption while
setting up the relations: that the recursive call in the body of
\verb"listfact" refers to the same predicate call, so we can set up a
recurrence equation. This fact is implicitly assumed in Hindley-Milner
type systems.  But in logic programming it is usual for a predicate to
be called with different patterns (such as different
modes). Fortunately, the \ciaopp\ framework allows multivariance
(support for different call patterns of the same predicate).  For the
analysis to handle it, we cannot just add calls with the bare name of
the predicate, because it will conflate all the existing versions. The
solution that we propose adds a new component to the abstract element:
a random name given to the specific instance of the predicate we are
analyzing, that is generated in the $\lambda_{call}$ to
$\beta_{entry}$. Then, in the widening step, all different versions of
the same predicate name are conflated.

\

Even though the analysis works with relations, these are not as useful
as functions defined without recursion or calls to other
functions. First of all, developers will get a better idea of the
sizes if presented in such a closed form. Second, functions are
amenable to comparison as outlined in~\cite{resource-verif-iclp2010},
which is essential for example in resource usage verification. There
are several software packages that are able to get bounds for
recurrence equations: computer algebra systems, such as Mathematica
(which has been integrated to get a fully automated analysis) or
Maxima; and specialized solvers such as PURRS~\cite{BagnaraPZZ05} or
PUBS \cite{conf-vmcai-AlbertGM11}. In our implementation we apply this
overapproximation operator after each widening step.  For our example,
the final abstract substitution is:
\begin{small}
$$\lambda'_1 \nabla \lambda'_2 =
\left\langle
\begin{array}{c}
(s_{L}, s_{U}), \{ (n_{L}, n_{U}) \}, \texttt{false}, \{ \alpha_1, \beta_1 \geq 0 \}, \\
\left\{ (s_L, s_U) \lessgtr (1,1), (n_L, n_U) \lessgtr (\alpha_1 \gamma_1, \beta_1 \delta_1) \right\},
\texttt{not\_fails}, \texttt{is\_det}
\end{array}
\right\rangle$$
\end{small}

\begin{table}[ht]
\caption{Experimental results.}\label{expresults}
\vspace{-0.5cm}
\begin{center}
\begin{tabular}{|l|ccc|ccccc|}
\hline
\hline
\textbf{\emph{Program}}
& \multicolumn{3}{c|}{\textbf{\emph{Resource Analysis (LB)}}}
& \multicolumn{5}{c|}{\textbf{\emph{Resource Analysis (UB)}}}
\\
& \emph{New} & \multicolumn{2}{c|}{\emph{Previous}} 
& \emph{New} & \multicolumn{2}{c}{\emph{Previous}} 
& \multicolumn{2}{c|}{\emph{RAML}} \\
\hline \hline
\texttt{append}
& $\alpha$
& $\alpha$
& = 
& $\beta$
& $\beta$
& =
& $\beta$
& = \\
\texttt{appendAll2}
& $a_1a_2a_3$
& $a_1$
& +
& $b_1b_2b_3$
& $\infty$
& +
& $b_1b_2b_3$
& = \\
\texttt{coupled}
& $\mu$
& $0$
& +
& $\nu$
& $\infty$
& +
& $\nu$
& = \\
\texttt{dyade}
& $\alpha_1\alpha_2$
& $\alpha_1\alpha_2$
& = 
& $\beta_1\beta_2$
& $\beta_1\beta_2$
& =
& $\beta_1\beta_2$
& = \\
\texttt{erathos}
& $\alpha$
& $\alpha$
& = 
& $\beta^2$
& $\beta^2$
& =
& $\beta^2$
& = \\
\texttt{fib}
& $\phi^\mu$
& $\phi^\mu$
& = 
& $\phi^\nu$
& $\phi^\nu$
& =
& infeasible
& + \\
\texttt{hanoi}
& $1$
& $0$
& +
& $2^\nu$
& $\infty$
& +
& infeasible
& + \\
\texttt{isort}
& $\alpha^2$
& $\alpha^2$
& = 
& $\beta^2$
& $\beta^2$
& =
& $\beta^2$
& = \\
\texttt{isortlist}
& $a_1^2$
& $a_1^2$
& = 
& $b_1^2b_2$
& $\infty$
& +
& $b_1^2b_2$
& = \\
\texttt{listfact}
& $\alpha\gamma$
& $\alpha$
& +
& $\beta\delta$
& $\infty$
& +
& unknown
& ? \\
\texttt{listnum}
& $\mu$
& $\mu$
& = 
& $\nu$
& $\nu$
& =
& unknown
& ? \\
\texttt{minsort}
& $\alpha^2$
& $\alpha$
& +
& $\beta^2$
& $\beta^2$
& =
& $\beta^2$
& = \\
\texttt{nub}
& $a_1$
& $a_1$
& = 
& $b_1^2b_2$
& $\infty$
& +
& $b_1^2b_2$
& = \\
\texttt{partition}
& $\alpha$
& $\alpha$
& =
& $\beta$
& $\beta$
& =
& $\beta$
& = \\
\texttt{zip3}
& $\min(\alpha_i)$
& $0$
& +
& $\min(\beta_i)$
& $\infty$
& +
& $\beta_3$
& + \\
\hline
\end{tabular}
\end{center}
\vspace{-0.7cm}
\end{table}

\vspace{-0.3cm}
\section{Experimental results}
\label{sec:results}

We have constructed a prototype implementation in \ciao{} by defining
the abstract operations for sized type and resource analysis that we
have described in this paper and plugging them into \ciaopp's PLAI
implementation. Our objective is to assess the gains in precision in
resource consumption analysis.

Table~\ref{expresults} shows the results of the comparison between the
new lower (\textbf{\emph{LB}}) and upper bound (\textbf{\emph{UB}})
resource analyses implemented in \ciaopp{}, which also use the new
size analysis (columns \emph{New}), and the previous resource analyses
in \ciaopp{}~\cite{caslog,low-bounds-ilps97,resource-iclp07} (columns
\emph{Previous}). We also compare (for upper bounds) with
\emph{RAML}'s analysis~\cite{DBLP:journals/toplas/0002AH12} (column
\emph{RAML}).

Although the new resource analysis and the previous one infer concrete
resource usage bound functions (as the ones
in~\cite{resource-iclp07}), for the sake of conciseness and to make
the comparison with RAML meaningful, Table~\ref{expresults} only shows
the complexity orders of such functions, e.g., if the analysis infers
the resource usage bound function $\Phi$, and $\Phi \in
\bigcirc(\Psi)$, Table~\ref{expresults} shows $\Psi$. The parameters
of such functions are (lower or upper) bounds on input data sizes. The
symbols used to name such parameters have been chosen assuming that
lists of numbers $L_i$ have size
$ln^{(\alpha_i,\beta_i)}(n^{(\gamma_i,\delta_i)})$, lists of lists of
lists of numbers have size
$llln^{(a_1,b_1)}(lln^{(a_2,b_2)}(ln^{(a_3,b_3)}(n^{(a_4,b_4)})))$,
and numbers have size $n^{(\mu,\nu)}$.  Table~\ref{expresults} also
includes columns with symbols summarizing whether the new \ciaopp\
resource analysis improves on the previous one and/or \emph{RAML}'s:
$+$ (resp. $-$) indicates more (resp.\ less) precise bounds, and $=$
the same bounds.  The new size analysis improves on \ciaopp's previous
resource analysis in most cases. Moreover, RAML can only infer
polynomial costs, while our approach is able to infer other types of
cost functions, as is shown for the divide-and-conquer benchmarks
\texttt{hanoi} and \texttt{fib}, which represent a large and common
class of programs. For predicates with polynomial cost, we get equal
or better results than RAML.

\section{Related work}
\label{sec:related}

Several other analyses for resources have been proposed in the
literature. Some of them just focus on one particular resource
(usually execution time or heap consumption), but it seems clear that
those analyses could be generalized.

We already mentioned RAML~\cite{DBLP:journals/toplas/0002AH12} in
Section~\ref{sec:results}. Their approach differs from ours in the
theoretical framework being used: RAML uses a type and effect system,
whereas our system uses abstract interpretation. Another important
difference is the use of polynomials in RAML, which allows a complete
method of resolution but limits the type of closed forms that can be
analyzed. In contrast, we use recurrence equations, which have no
complete decision procedure, but encompass a much larger class of
functions. Type systems are also used to guide inference in
\cite{grobauer01cost} and~\cite{igarashi02resource}.

In~\cite{nielson02automatic}, the authors use sparsity information to
infer asymptotic complexities. In contrast, we only get closed
forms. Similarly to \ciaopp's previous analysis, the approach
of~\cite{conf-vmcai-AlbertGM11} applies the recurrence equation method
directly (i.e., not within an abstract interpretation framework).
\cite{Rosendahl89} shows a complexity analysis based on abstract
interpretation over a step-counting version of functional programs.
\cite{DBLP:conf/ppdp/GieslSSEF12} uses symbolic evaluation graphs to
derive termination and complexity properties of logic programs.

\section{Conclusions and Future Work}

In this paper we have presented a new formulation of resource analysis
as a domain within abstract interpretation and which uses as input
information the sized types that we developed
in~\cite{sized-types-iclp2013}. We have seen how this approach offers
benefits both in the quality of the bounds inferred by the analysis,
and in the ease of implementation and integration within a framework
such as PLAI/\ciaopp.

In the future, we would like to study the generalization of this
framework to different behaviors regarding aggregation. For example,
when running tasks in parallel, the total time is basically the
maximum of both tasks, but memory usage is bounded by the sum of them.
Another future direction is the use of more ancillary analyses to
obtain more precise results. Also, since we use sized types as a
basis, any new research that improves such analysis will directly
benefit the resource analysis.  Finally, another planned enhancement
is the use of mutual exclusion analysis (already present in \ciaopp)
to aggregate recurrence equations in a better way.

\bibliographystyle{plain}

\end{document}